# Visualizing interaction-driven restructuring of quantum Hall edge states

Jiachen Yu*,[1,2], Haotan Han*,[1,2], Kristina G. Wolinski*,[1,2], Ruihua Fan[3], Amir S. Mohammadi[3], Tianle Wang[3], Taige Wang[3], Liam Cohen[4], Kenji Watanabe[5], Takashi Taniguchi[6], Andrea F. Young[4], Michael P. Zaletel[3], Ali Yazdani[†,1,2]

* These authors contributed equally to this work.

† Correspondence to: yazdani@princeton.edu

[1] *Joseph Henry Laboratories, Princeton University, Princeton, NJ 08544, USA*

[2] *Department of Physics, Princeton University, Princeton, NJ 08544, USA*

[3] *Department of Physics, University of California at Berkeley, Berkeley, CA 94720, USA*

[4] *Department of Physics, University of California at Santa Barbara, Santa Barbara, CA 93106, USA*

[5] *International Center for Materials Nanoarchitectonics,*

*National Institute for Materials Science, 1-1 Namiki, Tsukuba 305-0044, Japan*

[6] *Research Center for Functional Materials, National Institute for Materials Science, 1-1 Namiki, Tsukuba 305-0044, Japan*

**Many topological phases host gapless boundary modes that can be dramatically modified by electronic interactions. Even for the long-studied edge modes of quantum Hall phases[1,2], forming at the boundaries of two-dimensional (2D) electron systems, the nature of such interaction-induced changes has been elusive. Despite advances made using local probes[3–13], key experimental challenges persist: the lack of direct information about the internal structure of edge states on microscopic scales, and complications from edge disorder. Here, we use scanning tunneling microscopy (STM) to image pristine electrostatically defined quantum Hall edge states in graphene with high spatial resolution and demonstrate how correlations dictate the structures of edge channels on both magnetic and atomic length scales. For integer quantum Hall states in the zeroth Landau level, we show that interactions renormalize the edge velocity, dictate the spatial profile for copropagating modes, and induce unexpected edge valley polarization that differ from those of the bulk. While some of our findings can be understood by mean-field theory, others show breakdown of this picture,**

**highlighting the roles of edge fluctuations and inter-channel couplings. We also extend our measurements to spatially resolve the edge state of fractional quantum Hall phases and detect spectroscopic signatures of interactions in this chiral Luttinger liquid. Our study establishes STM as a promising tool for exploring edge physics of the rapidly expanding 2D topological phases, including newly realized fractional Chern insulators.**

A hallmark of topological phases is the presence of gapless edge modes. In the paradigmatic quantum Hall (QH) systems, they manifest as ballistic, chiral one-dimensional channels hosting gapless charge excitations[1]. Their existence and number are topologically protected and insensitive to local details, underpinning the universality of quantized Hall conductance[2]. However, this canonical picture breaks down when the edge confining potential is softened beyond a critical threshold, leading to a quantum phase transition at the edge known as the edge reconstruction[14,15]. Such reconstruction can occur in both the integer[14–19] and fractional QH regimes[20–25], introducing rich internal edge structures and giving rise to unusual edge excitations[24,25]. A central puzzle is how the edge modes are spatially distributed. Competing reconstruction scenarios predict patterns that vary at the magnetic length scale ($l_B$), making them difficult to resolve with conventional techniques like electrical transport. Even less is known about the edges of correlated ground states, such as quantum Hall ferromagnets (QHFM) and fractional quantum Hall (FQH) phases, where more complex edge modes emerge that can be understood only by probing on the microscopic scales.

Local scanning probe experiments[3–13] provide a promising approach for uncovering the edge state structure and their reconstruction. However, a direct visualization of the internal structure of QH edge states has remained elusive, hindered by disorder at the physical boundaries, or by insufficient sub-$l_B$ spatial resolution. Recent advancements in probing ultraclean graphene QH systems with scanning tunneling microscopy and spectroscopy (STM/S)[11,26–30] present an opportunity to address these challenges. Notably, substantial progress has been made in STM/S imaging of edge states at the graphene crystalline boundaries on graphite[9] and hexagonal boron nitride (hBN)[10], most relevant to our work on a p-n lateral interface[11]. Strong electronic interactions in graphene's zeroth Landau level (zLL) induce spin- and valley-ordered ground states[26,30,31], enriching edge reconstruction mechanisms with symmetry-breaking[32–34] and enabling visualization of correlation on the edge via broken symmetries.

Here we leverage the spatial resolution of STS to study electrostatically defined QH and FQH edge states in graphene, enabled by latest development in van der Waals device fabrication. Our approach minimizes edge disorder and allows control over the interplay between edge potential and other energy scales. Our results demonstrate that edge states can exhibit behavior not captured by the simple edge reconstruction models. Understanding this behavior is critical as edge states play an increasingly important role in various settings, such as probing the bulk topological order and in the recent anyon interferometry[35–38] and time-domain braiding[39,40] experiments.

**STM imaging of gate defined edge**

In this work, we use a recently developed local anodic oxidation technique (see Methods) to fabricate a graphene device with electrostatically defined edge (Fig. 1a,b). It has a global back gate (BG) and a patterned gate (PG) isolated by hBN to define a lateral interface. Gate voltages $V_{PG}$ and $V_{BG}$ independently control electron densities - and therefore Landau level (LL) filling factors - on PG and BG sides of the interface, denoted by $v_P$ and $v_B$.

We confirm the high quality of our gate-defined edge and electrical control of its potential via STS potential sensing[41] (Extended Data Figs. 1a,b). The intrinsic potential variation is a few mVs (Extended Data Fig. 1c, measured in the same region as in Fig. 1b), small compared with all relevant energy scales, e.g. gate-induced potential and electronic interactions. Tunability of the gate-induced potential is demonstrated by potential sensing with an imposed potential across the edge (Extended Data Fig. 1d, Supplementary Information Section 2).

Figure 1c shows representative gate-dependent spectra taken in the PG controlled region, exhibiting all features reported previously[26,27] (BG region shows similar features). An important spectroscopic feature for tunneling into QH states is the 'Haldane sashes' (black dashed square)[27], which appear only when the STM tip minimally perturbs the sample. We ensure this condition is always satisfied to minimize tip-induced influence on the edge states. The non-invasiveness of our experimental approach has been established by previous studies[26–29].

Before discussing interaction effects on QH edge modes, we show that for higher LLs ($N \neq 0$), single particle physics dominates. Figure 1d shows STS along a linear trajectory across the gate defined edge, with $v_B$ = -6 and $v_P$ = -2 (single-particle gaps). The $N$ = -1 LL disperses with distance ($x$) strongly in the incompressible bulk, reflecting LL bending by edge potential. The splitting in the $N$ = -1 and $N$ = -2 LLs in these dispersing segments matches the reported 'branching' behavior[11] and can be attributed to the massless Dirac LL wavefunction structure under confining potential. In contrast, near the edge, the $N$ = -1 LL is pinned to zero bias ($V_B$ = 0) over ~100 nm. This can be understood within the electrostatic-driven edge reconstruction model[14] in the soft potential limit, where the interplay of cyclotron energy and electrostatic potential creates an extended region where the LL is partially filled and pinned to $E_F$ (a compressible strip). From the slope of dispersing LLs, we estimate the edge potential energy scale $E_V = e\partial\phi/\partial x \cdot l_B \sim 5$ meV at $B$ = 6 T ($l_B$ = 10 nm), smaller than the cyclotron gaps, placing the experiment deep in the soft-potential limit. The absence of splitting in this compressible strip as the 4-fold degenerate $N$ = -1 LL intersects $E_F$ indicates that single-particle energetics governs the edge physics in higher LLs.

**Correlation-driven edge reconstruction**

Electronic correlations dominate when we set both $v_B$ and $v_P$ to the 4-fold degenerate zLL manifold, where QH ferromagnetic and FQH phases emerge. Figure 2 shows STS measured along the same trajectory as Fig. 1d with $v_B$ = -2 and $v_P$ = -1 (Fig. 2a), $v_P$ = 0 (Fig. 2b), and $v_P$ = 2 (Fig. 2c). Strikingly, one or multiple narrow peaks in d$I$/d$V$ occur at $V_B$ = 0 (Figs. 2d-f), signifying discretized edge channels and contrasting the wide compressible strip above. The number of zero bias peaks (ZBPs) satisfies $n$ = |$v_B$ - $v_P$|, experimentally demonstrating bulk-edge correspondence[2]. Gaussian fitting of ZBPs (Figs. 2d-f) yields channel widths of 13 - 15 nm, on the order of $l_B$. Similar spectral structure to Fig. 2c has been proposed to explain interferometric results[42]. We further investigate the spatial structure of these edge states by STS imaging, shown in Figs.

2h-j. These images are taken on a flipped L-shaped edge (Fig. 1b), where the equipotential contour turns 90 degrees (see Extended Data Fig. 1d, origins coincide). They show that the channels follow the local tangential direction of the potential gradient and remain well separated, showing a generic reconstruction pattern insensitive to small fluctuations of the underlying potential. These images constitute the first example of resolving channel splitting structure within a multi-channel QH edge.

Whereas the resolved edge structure qualitatively aligns with expectations, quantitative analysis shows an unexpected upward renormalization of edge velocity. For a single channel, we extract the $V_B$-dependent spatial displacement of the d$I$/d$V$ peak along the edge-normal direction ($\Delta x$) relative to $V_B = 0$ (Fig. 2k). $V_B$ varies quasi-linearly with $\Delta x$, the slope of which gives a spectroscopic determination of the edge velocity. The fitting window is chosen to capture the linear regime near $V_B = 0$ following standard definition of edge velocity. This measurement probes velocity equivalently to transport because they are both sensitive to excitations close to the $E_F$, but with the additional benefit of obtaining channel-resolved information in multi-channel situations. The edge mode velocities are ~ $10^5$ m/s for different ($v_B$, $v_P$), exceeding expectations from the steepness of the gate-induced confining potential (estimated from LL slopes away from $V_B = 0$) by ~4 times.

We perform Hartree-Fock (HF) calculations with a realistic edge potential profile and interaction parameters matching our experiment (see Methods and Supplementary Information Section 3 and 4). The obtained local density of states (LDOS) across the edge (Fig. 2g) shows remarkable qualitative consistency with experiment. Edge velocity can be analogously extracted and shows quantitative agreement for ($v_B$, $v_P$) = (-2, -1) ($\approx 1.3 \times 10^5$ m/s, see Supplementary Information Section 5). However, predictions for ($v_B$, $v_P$) = (-2, 0) deviate notably, which we attribute to the neglected inter-channel couplings that can renormalize edge velocities.

**Edge wavefunctions and broken symmetry**

Previous studies have not been able to directly probe isospin order of the edge states. The atomic resolution of STM/S allows us to directly visualize broken isospin symmetries in the zLL, because valley and sublattice polarization are directly linked in the zLL wavefunctions[26,27,30]. Specifically, by performing atomic scale d$I$/d$V$ mapping of the edge states and analyzing their fast Fourier transformations (FFT), we extract the isospin orientation, parametrized by polar ($\theta$) and azimuthal ($\varphi$) angles: $|\psi\rangle = \cos(\frac{\theta}{2})|K\rangle + \sin\left(\frac{\theta}{2}\right)e^{i\phi}|K'\rangle$, and the valley polarization $Z = \cos\theta$ [26,27] (see Extended Data Fig. 2, Supplementary Information Section). Applying this protocol to edge states in Figs. 2h-j, whose underlying wavefunctions belong to the zLL manifold, thus provides a long-sought direct measure of the edge isospin order. Figure 3a shows a d$I$/d$V$ map ($V_B$ = -1 mV) within the edge state in Fig. 2h, clearly showing that the wavefunction is sublattice-polarized, indicating valley polarization, with $Z$ = 0.45. This behavior can be qualitatively captured by the H-F (Fig. 3c, left), which predicts that $Z(x)$ continuously interpolates between the nearby bulk values ($Z$ = 0 for $v_B$ = -2, $Z \approx 0.75$ for $v_P$ = -1, see Supplementary Information Section 4.2), resulting in an intermediate $Z$ at the edge.

In contrast to the single-channel case, when multiple edge channels are present, breakdown of the mean-field reconstruction becomes evident, as exemplified ($v_B$, $v_P$) = (-2, 0). Figure 3b shows the wavefunction imaged on the left and right channels in Fig. 2i. Both exhibit Kekulé distortion patterns, indicative of intervalley coherence[26,27,30] (IVC, see Extended Data Fig. 2 for FFT). This contradicts the mean-field expectation that a $v$ = -1 incompressible strip develops between the two edge channels, which maps the left channel to that observed in the ($v_B$, $v_P$) = (-2, -1) case. The spectroscopy gap in between two channels is significantly softened compared to the bulk $v$ = -1 gap (Fig. 2b), further distinguishing it from a bulk $v$ = -1 state. Thus, the two edge channels plus the intermediate region constitute a reconstructed system whose properties differ fundamentally from existing mean-field descriptions. This is further supported by the isospin order measured for both channels (Fig. 3b), which surprisingly show finite and opposite $Z$ values of comparable magnitude. This subtle discrepancy in $Z$ can be visualized directly by filtered inverse FFT to Fig. 3b retaining only the Bragg peaks to highlight valley polarization contrasts (Extended Data Fig. 3). Mapping $Z$ along the two edge channels (Fig. 3d) shows that this opposite-sign $Z$ correlation is a robust feature persisting along the edge, which clearly contradicts the H-F prediction that the two channels share the same *Z* sign (Fig. 3c, right). We posit beyond mean-field fluctuations soften the charge gap of the intermediate incompressible strip and perturb the edge state isospin. Inter-channel coupling can also generate unexpected correlations between co-propagating channels[43]. Together they highlight the rich correlation effects on edge structures long overlooked in the most studied QH systems.

**Signature of fractional edge state**

We further explore FQH edge states within the zLL. Unlike integer states, their low-energy physics follows chiral Luttinger liquid (CLL) theory[44], whose collective charge excitations are orthogonal to the single-particle eigenstates, leading to a universal power law suppression of single-electron tunneling near $E_F$[45]. This suppression invalidates the detection scheme described above, but leaves a distinct signature on the functional dependence of $I(V_B)$. Earlier global tunneling measurements on cleaved edge[45,46] or in point contact[47] devices have used this property to diagnose CLL, whereas our setup accesses the same information microscopically.

Figure 4a shows $|I|$ along a line across the edge with $(v_B, v_P) = (1, 2/3)$, where bulk-edge correspondence predicts one fractional edge channel. This choice of $v_P$ maximizes FQH gap[26,27] and enhances the spectroscopic contrast between compressible and incompressible regions (see Fig. S8 for the bulk FQH gap and determination of $v_P$). The bulk on opposite ends of this trajectory are gapped, but the middle region exhibits a distinctive behavior: the onset of $|I|$ occurs at significantly lower $|V_B|$, suggestive of gapless excitations; yet strong suppression of $|I|$ close to $V_B = 0$ contrasts it with the integer edge situations. We quantify this by comparing the spectroscopically determined tunneling gap $\Delta$ (blue curve in Fig. 4b, see Methods) with that of a compressible bulk state, the latter characterized by a soft Coulomb gap ($\Delta_C$, gray line in Fig 4b). For $\Delta > \Delta_C$, the system is incompressible, whereas $\Delta < \Delta_C$ signals gapless edge excitations with reduced Coulomb suppression, distinct from compressible bulk excitations. From this we identify an edge region of width ~ $2l_B$ where $\Delta < \Delta_C$, hinting at the presence of fractional edge excitation.

$I$-$V_B$ characteristics for tunneling into edge states provide a quantitative method to characterize the edge excitations. Figures 4c,d show representative $I$-$V_B$ measured for integer and fractional edge states, and our fitting attempts to the predicted power-law form $I = kV_B^\alpha$. For IQH edge states, the simplest model predicts a linear ($\alpha = 1$) $I$-$V_B$, but we find a deviation from this prediction both at low biases, where $I$-$V_B$ appears non-power-law, as well as a larger power-law exponent, $\alpha = 1.89 \pm 0.05$, at high biases. For FQH states, we obtain $\alpha = 3.20 \pm 0.03$, closer to but larger than the predicted universal value of $\alpha = 3$. The edge state's $I$-$V_B$ also shows asymmetric bias dependence for both IQH and FQH (Supplementary Information Section 8). Despite the deviation from predictions, the $I$-$V_B$ at the edges for both IQH and FQH are distinct from tunneling into the compressible bulk states, the latter of which features a large $\Delta_C$ near $V_B = 0$ (power-law fits yield significantly larger $\alpha = 4 - 5$ for $V_B > 0$).

We further establish a concrete link between the representative $I$-$V_B$ in Fig. 4e and fractional edge excitation by mapping $\alpha$ along the edge. Figure 4e shows $\alpha$ extracted at each spatial point from fitting the $V_B > 0$ segment of $I(V_B)$ over a rectangular region along the edge (inset), with $(v_B, v_P) = (1, 2/3)$. Spatial points where $|I|$ signal is too weak to perform reliable fits are excluded, such as those in the incompressible state (empty pixels, see Supplementary Information Section 8 for protocols). We resolve a ~ 20 nm strip adjacent to the $v_B = 1$ bulk, where $\alpha$ lies predominantly within (2.9, 3.6). In marked contrast, by setting $v_P = 0.73 > 2/3$ (compressible state) to suppress the fractional edge state (Fig. 4f), we observe $\alpha > 3.6$ throughout the entire compressible region up to the $v_B = 1$ bulk boundary. The smaller $\alpha$ indicative of edge state is cross-validated by performing the same analysis on an IQH edge (see Supplementary Information Section 8.3, Fig. S12). Overall, while the $I$-$V_B$ characteristic and its spatial dependence establish our ability to probe edge states with STM, the deviation of our results from various predictions raises questions as to how our experiment probes such states. Lack of electronic momentum conservation and other differences between STM tunneling and experiments involving cleaved edge[45] or point contact[47] tunneling may have to be theoretically examined to understand our experimental findings.

**Outlook**

We provide a detailed microscopic view of how interactions modify some of the most-studied topological edge states, previously inaccessible experimentally. The new physical insights motivate future works to address the discrepancies between our experiment and existing theories. The platform is highly tunable (Extended Data Fig. 4), enabling in-depth, quantitative studies of potential-tuned edge configurations. Extending to lower temperatures presents the intriguing possibility of addressing long-standing debates concerning the pattern of co- and counter-propagating modes in hole-conjugate[20,21,23] and non-abelian[24] FQH states. This understanding is foundational in the development of quantum devices that leverage edge states for processing quantum information[35–38,48]. The approach can also disentangle the edge structure where integer and fractional modes coexist, (e.g. even-denominator states of bilayer graphene[28,49]), and is applicable to a range of 2D topological states, including quantum anomalous Hall and fractional Chern insulators[50–53], whose edge excitations constitute a rich yet largely unexplored playground.

# References

1. Halperin, B. I. Quantized Hall conductance, current-carrying edge states, and the existence of extended states in a two-dimensional disordered potential. *Phys. Rev. B* **25**, 2185–2190 (1982).

2. Hatsugai, Y. Chern number and edge states in the integer quantum Hall effect. *Phys. Rev. Lett.* **71**, 3697–3700 (1993).

3. Yacoby, A., Hess, H. F., Fulton, T. A., Pfeiffer, L. N. & West, K. W. Electrical imaging of the quantum Hall state. *Solid State Commun.* **111**, 1–13 (1999).

4. Paradiso, N. *et al.* Imaging Fractional Incompressible Stripes in Integer Quantum Hall Systems. *Phys. Rev. Lett.* **108**, 246801 (2012).

5. Pascher, N. *et al.* Imaging the Conductance of Integer and Fractional Quantum Hall Edge States. *Phys. Rev. X* **4**, 011014 (2014).

6. Suddards, M. E., Baumgartner, A., Henini, M. & Mellor, C. J. Scanning capacitance imaging of compressible and incompressible quantum Hall effect edge strips. *New J. Phys.* **14**, 083015 (2012).

7. Cui, Y.-T. *et al.* Unconventional Correlation between Quantum Hall Transport Quantization and Bulk State Filling in Gated Graphene Devices. *Phys. Rev. Lett.* **117**, 186601 (2016).

8. Uri, A. *et al.* Nanoscale imaging of equilibrium quantum Hall edge currents and of the magnetic monopole response in graphene. *Nat. Phys.* **16**, 164–170 (2020).

9. Li, G., Luican-Mayer, A., Abanin, D., Levitov, L. & Andrei, E. Y. Evolution of Landau levels into edge states in graphene. *Nat. Commun.* **4**, 1744 (2013).
10

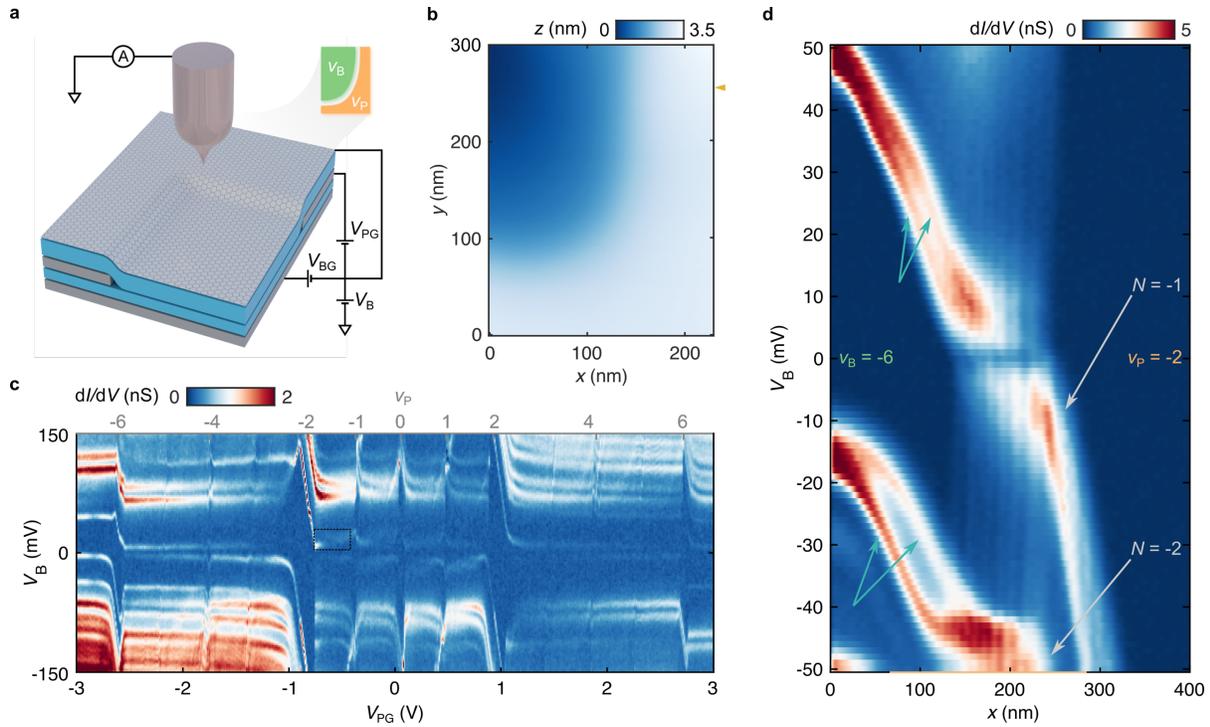

**Figure 1 | Electrostatically defined edge in graphene. a,** Schematic of the measurement setup. DC voltages applied to the patterned graphite gate ($V_{PG}$, middle layer) and global back gate ($V_{BG}$, bottom layer) independently tune the filling factor of two regions in monolayer graphene ($\nu_P$, yellow; $\nu_B$, green, see Methods). $V_B$ controls the tunneling bias between graphene and the tip, whereas tunneling current ($I$) and differential conductance ($dI/dV$) is measured from the tip. Gray and blue layers represent graphite and hexagonal boron nitride (hBN), respectively. **b,** Scanning tunneling microscopy (STM) measurement of the graphene topography ($V_B$ = 400 mV, $I$ = 50 pA). Back gate controlled region shows a ~3.5 nm height decrease, consistent with the thickness of the patterned graphite gate. **c,** Representative scanning tunneling spectroscopy (STS) as a function of $V_{PG}$ taken at a patterned gate controlled region. **d,** STS measured along a linear trajectory across the gate defined edge ($y$ = 271 nm, yellow triangle), with $\nu_B$ = -6 and $\nu_P$ = -2, respectively. Two pairs of teal arrows annotate the 'branching' feature[11] (see main text). Yellow line overlaid on the x-axis denotes the spatial extent of the scanned area in **b.**

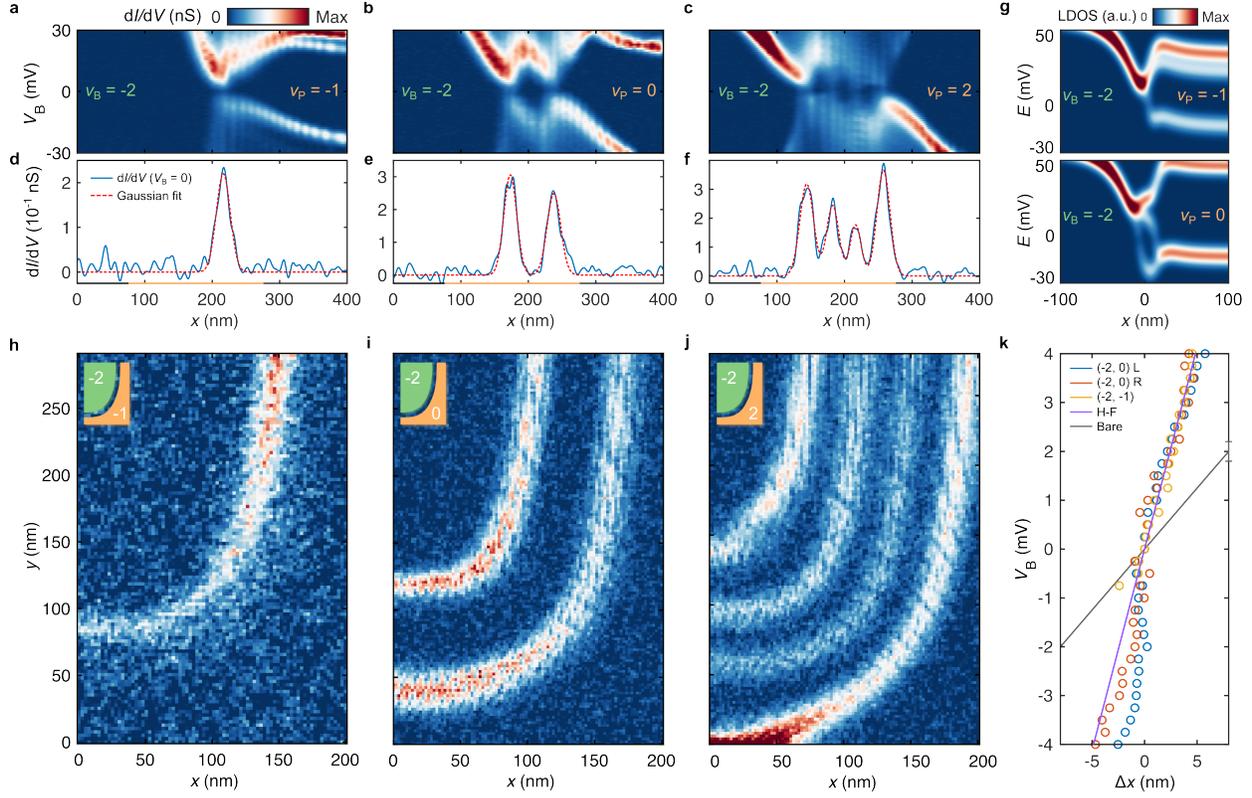

**Figure 2 | Imaging edge states and their reconstruction. a-c,** STS taken along a linear trajectory across the edge, with filling factors ($v_B$, $v_P$) = (-2, -1) (**a**), (-2, 0) (**b**), and (-2, 2) (**c**). **d-f,** d$I$/d$V$ at $V_B$ = 0, with filling factors ($v_B$, $v_P$) = (-2, -1) (**d**), (-2, 0) (**e**), and (-2, 2) (**f**). The number of zero bias peaks (ZBP) is consistent with the total change of filling factor across the edge. Gaussian fittings of the ZBPs yield peak widths ~ 13 - 15 nm, on the order of magnetic length $l_B$ = 10 nm. **h-j,** STS imaging taken at $V_B$ = 0, with filling factors ($v_B$, $v_P$) = (-2, -1) (**h**), (-2, 0) (**i**), and (-2, 2) (**j**). The origin of the coordinate is the same spatial points as that of Fig. 1b. (**a-f**) are taken at $y$ = 271 nm, and yellow lines overlaid on the $x$-axes denote the spatial extent of (**h-j**) in the horizontal direction. **g,** Hartree-Fock calculations of the local density of states as probed by STM, for ($v_B$, $v_P$) = (-2, -1) (upper panel) and (-2, 0) (lower panel). **k,** Energy-dependent d$I$/d$V$ peak position shifts relative to ZBP ($\Delta x$). Edge velocity can be estimated from the slopes of linear fits: (-2, -1) in **a**, 1.33×10$^5$ m/s; (-2, 0) left in **b**, 1.71×10$^5$ m/s; (-2, 0) right in **b**, 1.52×10$^5$ m/s. Purple line represents velocity estimated from HF calculations for ($v_B$, $v_P$) = (-2, -1) (≈1.3×10$^5$ m/s); gray line represents velocity estimated from bare potential, with error bar indicating its uncertainty.

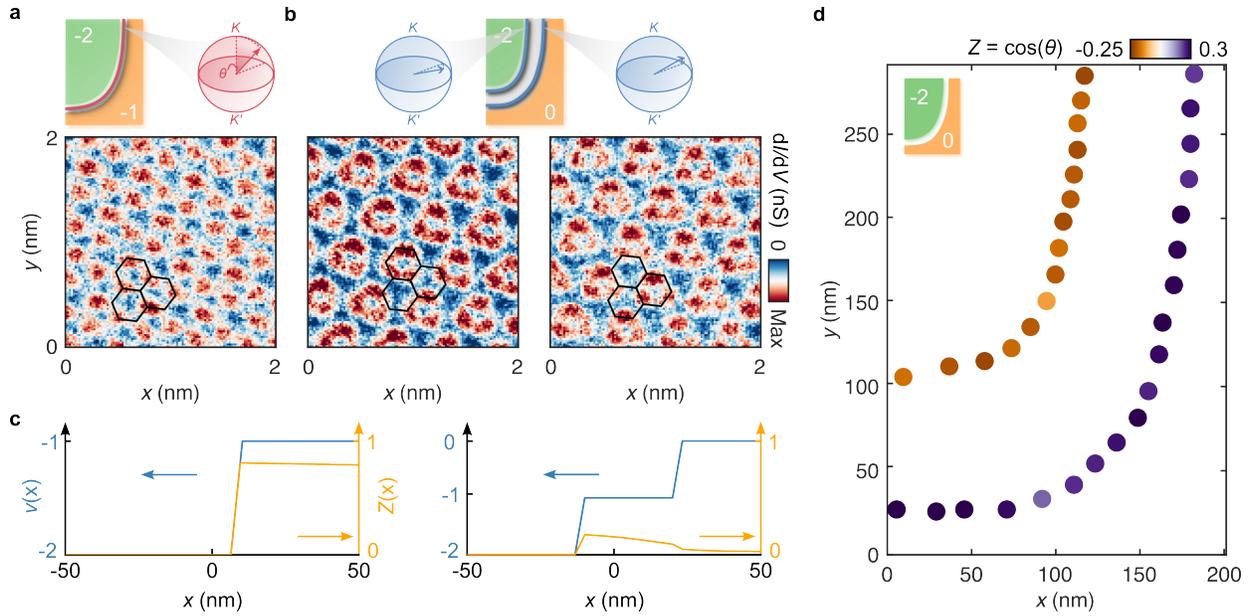

**Figure 3 | Imaging edge state wavefunctions and their valley isospin. a,** Atomic scale imaging of the edge state wavefunction at a representative point along the edge channel, with ($v_B$, $v_P$) = (-2, -1) at $V_B$ = -1 mV. The small negative bias ensures sufficient signal-to-noise ratio for FFT analysis (see Supplementary Information Section 6 for details of $V_B$ dependence). Overlaid black hexagons denote positions of carbon atoms in the graphene lattice. Upper panel, schematic illustration of the edge state valley order obtained from fast Fourier transform (FFT) analysis (see Supplementary Information Section 7). **b,** Similar representative images for both edge channels (left, left channel; right, right channel), taken with ($v_B$, $v_P$) = (-2, 0) at $V_B$ = 0. **c,** Hartree-Fock calculations of local filling factor $v(x)$ (blue) and valley polarization $Z(x)$ (yellow) for fillings (-2, -1) (left) and (-2, 0) (right). **d,** $Z$ measured along the two edge states of ($v_B$, $v_P$) = (-2, 0) (Fig. 2i). The image is obtained by performing atomic scale maps at the scattered points' locations along the two channels, and $Z$ is encoded as the colors of the scattered points (see Supplementary Information Section 7).

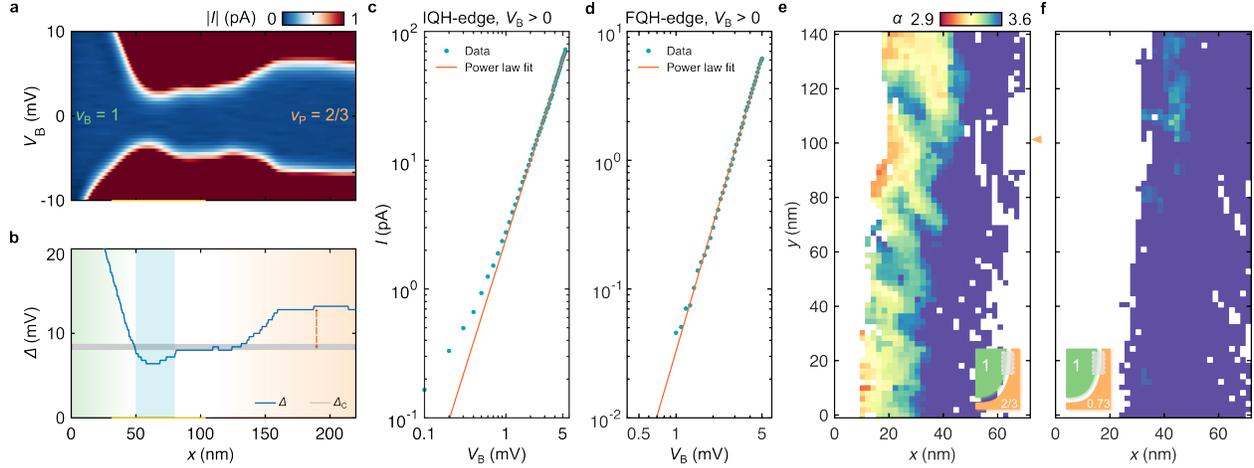

**Figure 4 | Spectroscopic signature of fractional edge mode. a,** Magnitude of tunneling current ($|I|$) measured along a linear trajectory across the edge, taken with ($v_B$, $v_P$) = (1, 2/3). **b,** Extracted spectroscopy gap $\Delta$ from **a** along the linear trajectory (see main text for extraction criteria). Gray horizontal line denotes the bulk Coulomb gap $\Delta_C$ [26] (see Supplementary Information Section 8 for details of the extraction procedure), whose thickness reflects uncertainty. Shaded green and yellow region represents areas in the incompressible states ($v_B$ = 1, green; $v_P$ = 2/3, yellow), determined by $\Delta > \Delta_C$. Dashed orange arrow denotes spectroscopic gap of the bulk $v_P$ = 2/3 state. Shaded cyan region represents an area where $\Delta < \Delta_C$. **c,d,** Power law fit ($I = kV_B^\alpha$) of the $I$-$V_B$ characteristic of the integer (**c**) and fractional (**d**) QH edge states (for $V_B > 0$), yielding $\alpha$ = 1.89 ± 0.05 (**c**) and $\alpha$ = 3.20 ± 0.03 (**d**). **d** is taken at $x$ = 60 nm in **b**. Datapoints below the measurement noise floor (40 fA) are removed. **e,f,** Power law fit parameter $\alpha$ imaged in a rectangular region along the edge, with $v_P$ = 2/3 (**e**) and in the compressible regime $v_P$ = 0.73 (**f**). Data is truncated for < 2.9 and > 3.6 in the colormaps of **e,f**. Empty pixels represent spatial points excluded due to small $|I|$ (max $|I|$ < 0.5 pA threshold or insufficient decade) that results in unreliable fit. Arrow in **e** indicates $y$ at which **a** is taken, whereas the spatial extent of **e,f** is denoted by the yellow line overlaid on the $x$-axis in **b**. Dashed rectangles in **e,f** insets sketch position of the imaged area on the edge. **a-f** measured with junction resistance of 12.5 MOhm.

## Methods

### Local anodic oxidation patterning of graphite gate

The patterned graphite gate (PG) is fabricated by performing local anodic oxidation[54,55] (LAO) in a Bruker Dimension Icon atomic force microscope (AFM). First, graphite is exfoliated onto a 90 nm $SiO_2$/Si wafer, which improves cutting efficiency. Then, a humid environment is established within the AFM by placing a beaker of 250 mL of deionized water on a hot plate at 120 °C, which is switched off by a bang-bang style humidity controller when the relative humidity reaches >50%. Etching is achieved by moving a conducting AFM tip along a thin graphite flake while an AC voltage (18V peak-to-peak amplitude, 200 kHz) is applied between the tip and flake. This catalyzes a local oxidation reaction and removes the flake material along the tip's path. Desired graphite shape can therefore be achieved by an appropriately designed tip trajectory. The gate used in this study contains four long (> 10 um) and wide (~ 1 um) trenches (Fig. S1) to aid with STM navigation. The edge state measurements are taken at the bottom corner of one of these trenches (Fig. S1). STM topography (Fig. 1b) of graphene on this region shows a smooth and monotonic ramp of 3.5 nm height, consistent with the thickness of the PG (Supplementary Information Section 1). This suggests that minimal oxidative residue from the graphite patterning process is picked up in the device fabrication process.

### Device fabrication

Monolayer graphene, graphite, and hexagonal boron nitride (hBN) flakes are exfoliated onto 285 nm $SiO_2$/Si wafers and characterized by optical microscopy. Their thickness and surface cleanliness are checked through AFM. The thickness of the graphite flake used as the PG is chosen to be 3.5 to 4 nm to ensure good metallic behavior. A top hBN of ≈ 52 nm thickness is selected to minimize gate screening of the Coulomb interaction in graphene, thereby enhancing the correlation gaps. The PG is etched with an AFM-based local anodic oxidation technique as detailed above. The selected van der Waals flakes are assembled into a heterostructure (Fig. 1a) by a polyvinyl-alcohol (PVA)-based dry transfer technique and dropped onto a Si/SiO$_2$ substrate with pre-patterned gold contacts, followed by extensive solvent cleaning procedure[26,27,41]. The completed device is wire bonded on a custom-built sample holder and annealed in ultrahigh vacuum (UHV) at 400 °C for 12 h to further remove surface polymer residue.

**STM measurements**

The experiment is done in a home-built UHV STM operating at electronic temperature $T$ = 1.4 K. All data shown are taken in an out-of-plane magnetic field $B$ = 6 T. The measurements are performed with a tungsten tip prepared on a Cu(111) single crystal. The specific tip preparation technique we adopt to minimize tip doping effect on graphene is elaborated in [27], and the influence of tip on the edge potential is further discussed in Supplementary Information Section 9. A capacitance navigation method is used to locate the sample[56].

In our experiments, the tip is grounded through the virtual ground of the transimpedance amplifier (NF SA606-S2), and bias voltage $V_B$ is applied to the monolayer graphene. DC voltages $V_{PG}$ and $V_{BG}$ are added onto $V_B$ and applied to the PG and BG, respectively. For spectroscopic measurements, AC excitation of frequency 697.77 Hz and amplitude 0.5 - 2 mV is added onto $V_B$ and both DC current and d$I$/d$V$ signals are simultaneously collected from the tip, where the latter is detected with standard lock-in techniques. For spectroscopy measurements with feedback loop off, junction resistance of 200 - 400 MOhm at $V_B$ = 400 mV was used for d$I$/dV, whereas for current measurement in the fractional regime the junction resistance is further reduced to 12.5 - 100 MOhm at $V_B$ = 400 mV to optimize signal to noise ratio (see Supplementary Information Section 8).

**Valley isospin extractions**

The degree of valley polarization, $Z$ = cos($\theta$), is extracted by taking the Fourier transform of the atomic-scale d$I$/d$V$ maps and analyzing the Fourier peaks[26,27]. Six peaks are present in all maps, corresponding to the graphene lattice (Extended Data

Fig. 2a). Three of these peaks, $G_1$, $G_2$, $G_3$, are used to calculate $Z = -\tan(\arg(n(G_1)n(G_2)n(G_3)/3))/3$, where $n$ is the complex Fourier amplitude. Some maps contain additional Fourier peaks (Extended Data Figs. 2b-e), indicative of the tripling of the unit cell that is characteristic of Kekulé distortion, and hence intervalley coherent order. Three of these additional peaks, $K_1$, $K_2$, $K_3$, are used to calculate the intervalley coherence phase $\phi$ as $\phi = \arg(n(K_1)n(K_2)n(K_3))/3$ (see Supplementary Information Section 7 for more details).

**Tunneling gap $\Delta$ extraction**

Tunneling gap $\Delta$ in the $I$-$V_B$ measurement (e.g. Fig. 4a) is extracted by the onset $V_B$ where $|I| = 0.5$ pA for both $V_B > 0$ and $V_B < 0$, namely $\Delta = V_B(I > 0.5 \text{ pA}) - V_B(I < -0.5 \text{ pA})$ (blue curve in Fig. 4b). The extracted $\Delta$ can be decomposed into a near-constant contribution from the soft Coulomb gap ($\Delta_C$) near $E_F$ and the additional energy cost of creating particle-hole excitations $\Delta_{ph}$, $\Delta = \Delta_C + \Delta_{ph}$ [27,28]. $\Delta_C$ can be measured within the bulk by setting $v_P = 0.73$ (a compressible state with $\Delta_{ph} = 0$, see Supplementary Information Section 8 for more details on the extraction). $\Delta_C$ is then used as a phenomenological benchmark to be compared with $\Delta$ for all spatial locations (Fig. 4b). For example, when $\Delta > \Delta_C$, the system is incompressible ($\Delta_{ph} > 0$).

**Hartree-Fock calculations**

There are four-flavor electrons in the zLL of the monolayer graphene. We model the interaction between electrons by an SU(4) symmetric screened Coulomb potential, plus an SU(4)-breaking short-range interaction. The Coulomb potential, giving the dominating energy scale, is determined directly from the actual electrostatics of the device. The short-range interaction, which involves four unknown parameters, is determined by matching the Hartree-Fock prediction on the spectrum and isospin order with the experimental results. Specifically, we use the data for the $v = 1$ bulk to fix parameters and verify its validity by complementary data (see Supplementary Information Section 4 for more details).

## Methods References

## Acknowledgements

We thank Ady Stern for helpful discussions. The work at Princeton is primary supported by the U.S. DOE-BES grant DE-FG02-07ER46419 and by the Gordon and Betty Moore Foundation's EPiQS initiative grant GBMF9469. Other support was provided by ONR grant N000142412471, NSF-MRSEC through the Princeton Center for Complex Materials grant NSFDMR-2011750, NSF grant DMR-2312311, NSF grant OMA-2326767, and the U.S. Army Research Office MURI project under grant number W911NF-21-2-0147. J.Y. is supported by the Princeton Materials Science Postdoctoral Fellowship. K.G.W. is supported by the Department of Defense (DoD) through the National Defense Science & Engineering Graduate (NDSEG) Fellowship Program. M.Z., R.F., A.S.M., Tianle Wang and Taige Wang are funded by the U.S. Department of Energy, Office of Science, Office of Basic Energy Sciences, Materials Sciences and Engineering Division under Contract No. DE-AC02-05-CH11231 (Theory of Materials program KC2301). R.F. is also supported by the Gordon and Betty Moore Foundation (Grant GBMF8688). Work at UCSB was supported primarily by the ONR under award N00014-23-1-2066 as well as by a Brown Investigator Award to A.F.Y.. K.W. and T.T. acknowledge support from the JSPS KAKENHI (grant nos. 21H05233 and 23H02052) and the World Premier International Research Center Initiative, MEXT, Japan.

## Author Contributions Statement

J.Y., H.H., K.G.W. and A.Y. conceived the experiment; J.Y., H.H., K.G.W. and A.Y. performed the STM measurements and analyzed the data; H.H. fabricated the device structure with help from J.Y., K.G.W. and L.C.; L.C. and A.F.Y. provided the patterned graphite gate; R.F., A.S.M., Tianle Wang, Taige Wang, and M.P.Z. carried out the theoretical calculations; K.W. and T.T. provided the h-BN crystals. All authors contributed to the writing of the paper.

**Data Availability Statement**

The data that support the findings of this study are available from the corresponding author upon reasonable request.

**Competing Interests Statement**

The authors declare no competing interests.

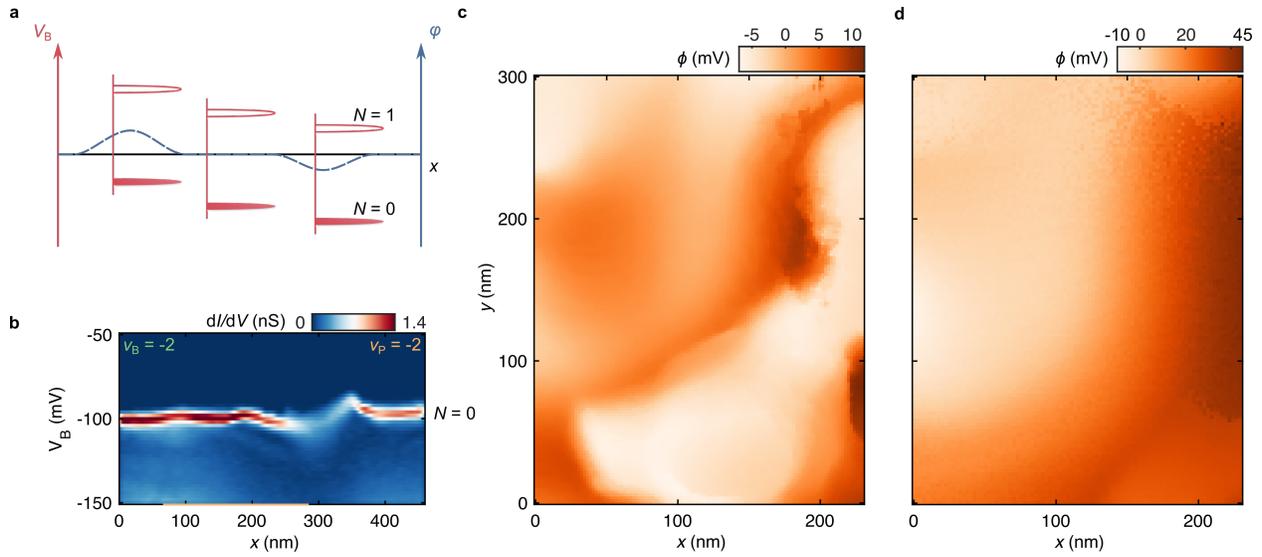

**Extended Data Figure 1 | Sensing and tuning edge potential. a,** Schematic illustration of the working principle of STS potential sensing. When graphene is highly incompressible (e.g. $v = 2$), variation in local potential $\varphi$ as felt by graphene (blue) shifts the energy of LLs (red), the latter of which is measured by the $V_B$ at which resonant tunneling to LLs occurs, i.e. peak in $dI/dV(V_B)$. **b,** STS measured along a linear trajectory across the gate defined edge with $v_B = v_P = 2$. $V_{PG}$ and $V_{BG}$ are adjusted within the gap to balance $\phi$ in the PG- and BG- controlled bulk to extract the intrinsic potential variations. **c,d,** $\phi$ extracted from the STS mapping near a flipped L-shaped edge of PG (same as Fig. 1b) with $v_B = v_P = 2$, where $V_{PG}$ and $V_{BG}$ are adjusted within the gap to balance (**c**) or offset (**d**) $\varphi$ across the edge. **b** is taken at $y = 300$ nm in **c,d**. Yellow line overlaid on the x-axis in **b** denotes the spatial extent of the scanned area in **c,d**.

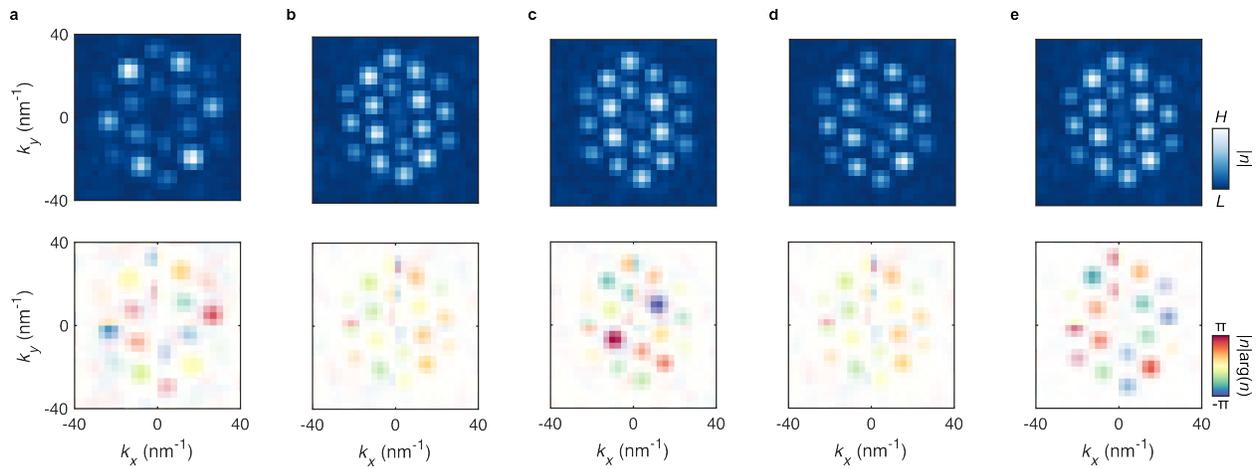

**Extended Data Figure 2 | Representative FFT images obtained from the atomic-scale STS maps of edge states. a-e,** Upper panels, modulus of the complex FFT amplitude $|n(\mathbf{k})|$; Lower panels, complex phase masked by the modulus, $|n(\mathbf{k})|\arg(n(\mathbf{k}))$ (see Methods and Supplementary Information Section 7 for details). Images are obtained from representative points on the edge channels in Fig. 2h (**a**), the two edge channels in Fig. 2i (left, **b;** right, **c**), and the two channels in Fig. S5a (left, **d;** right, **e**). $V_B = 0$ except for **a** where a small negative offset $V_B = -1$ mV is applied to maintain sufficient d$I$/d$V$ contrast for accurate FFT analysis. The six prominent peaks in the top panel of **a**, and the corresponding peaks at the same ($k_x$, $k_y$) in **b-e** are Bragg peaks of the graphene lattice. In **b-e** a new set of peaks at the Kekulé vectors are more prominent than the Bragg peaks, indicating strong intervalley coherence[27].

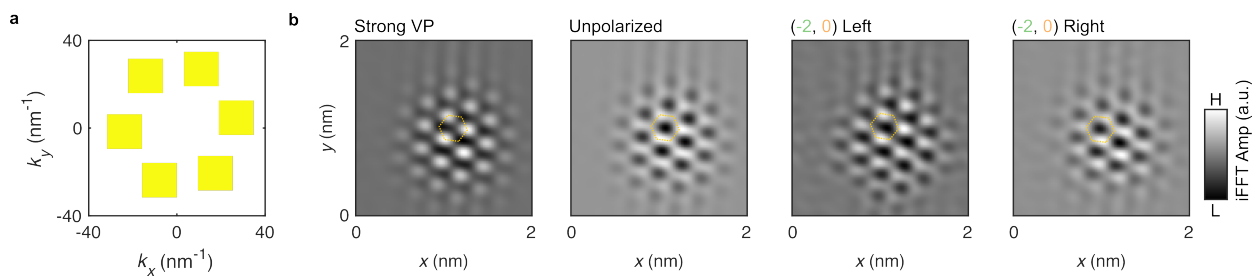

**Extended Data Figure 3 | Inverse Fast Fourier Transform (iFFT) analysis of valley polarization. a,** Fourier filter for retaining only the Bragg peaks in the FFT data. **b-e,** Representative examples of iFFT generated real-space intensity maps, for a strongly valley polarized state ($Z \sim 1$, **b**); graphene lattice ($Z = 0$, **c**); left (**d**) and right (**e**) panels in Fig. 3b. Partial valley polarization is visible in **d,e** as larger iFFT amplitude on one

sublattices. Yellow hexagons denote expected graphene atomic positions. Weaker contrast near the boundaries of **b-e** is a result of the Blackman window function applied to obtain the FFT peaks.

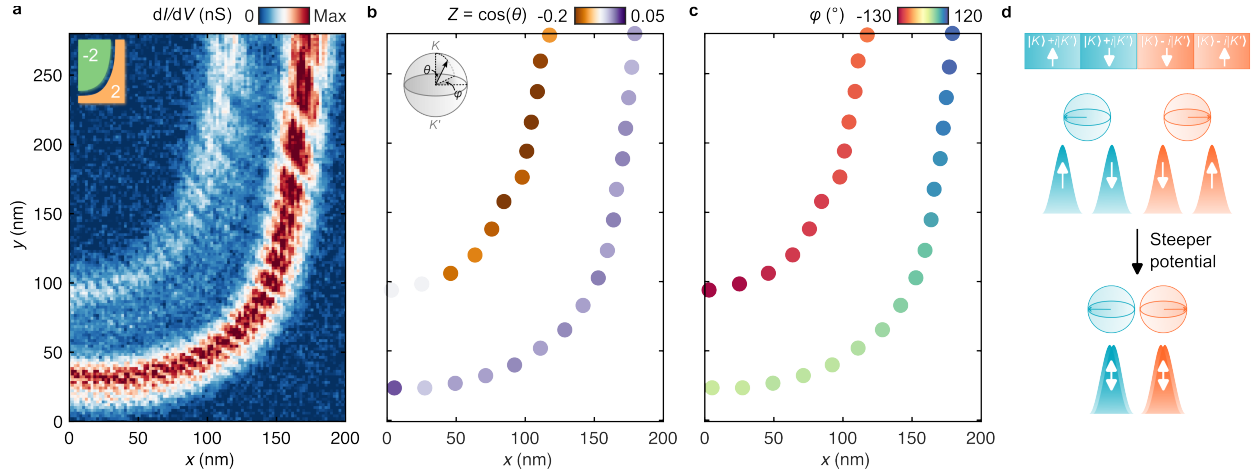

**Extended Data Figure 4 | Tuning the reconstruction of co-propagating edge states. a,** STS imaging of edge states with ($v_B$, $v_P$) = (-2, 2), but with a sharper edge potential compared to Fig. 2j. The four edge channels in Fig. 2j are 'squeezed' into two pairs of spatially overlapping edge channels. **b,c,** Valley polarization $Z = \cos\theta$ (**b**) and inter-valley coherence phase $\varphi$ (**c**) measured along the two edge channels in **a**. Inset in **b** illustrates valley Bloch sphere with $\theta$, $\varphi$ specified. **d,** cartoon illustration of the potential-tuned edge spin transition inferred from isospin measurements. In a smooth potential as in Fig. 2j, four channels acquire four distinct flavors (top panel). The spin anti-alignment within the left (cyan) and right (orange) pairs of edge states is inferred from the same isospin orientation within each pair and Pauli exclusion (middle panel). By increasing the steepness of the potential (bottom panel), each pair merges into a single channel, and retains its isospin orientation as they merge, therefore resulting in two spin-unpolarized channels. This demonstrates a potential-tuned spin phase transition at the edge[16,17].